\documentclass[10pt,twocolumn,english,prl,aps,floatfix,superscriptaddress]{revtex4}
\usepackage{babel}
\usepackage{verbatim}
\usepackage{float}
\usepackage{amsmath}
\usepackage{amssymb}
\usepackage{graphicx}
\usepackage{color}
\usepackage[pdfstartview={FitH},bookmarks=false]
 {hyperref}
\usepackage[T1]{fontenc}
\usepackage{bbold}

\makeatletter
\@ifundefined{textcolor}{}
{%
 \definecolor{BLACK}{gray}{0}
 \definecolor{WHITE}{gray}{1}
 \definecolor{RED}{rgb}{1,0,0}
 \definecolor{GREEN}{rgb}{0,1,0}
 \definecolor{BLUE}{rgb}{0,0,1}
 \definecolor{CYAN}{cmyk}{1,0,0,0}
 \definecolor{MAGENTA}{cmyk}{0,1,0,0}
 \definecolor{YELLOW}{cmyk}{0,0,1,0}
}

\begin{document}


\title{Coherence Based Characterization of Macroscopic Quantumness}

\author{Moein Naseri}
\affiliation{%
Department of Physics, Sharif University of Technology, Tehran, Iran
}%
\author{Sadegh Raeisi}%
 \email{sraeisi@sharif.edu}
\affiliation{%
Department of Physics, Sharif University of Technology, Tehran, Iran
}%

\begin{abstract}
One of the most elusive problems in quantum mechanics is the transition between classical and quantum physics. 
This problem can be traced back to the Schr\"{o}dinger's cat. 
A key element that lies at the center of this problem is the lack of a clear understanding and characterization of macroscopic quantum states. 
Our understanding of Macroscopic Quantumness relies on states such as the Greenberger-Horne-Zeilinger(GHZ) or the NOON state.  
Here we take a first principle approach to this problem. 
We start from coherence as the key quantity that captures the notion of quantumness and demand the quantumness to be collective and macroscopic. 
To this end, we introduce macroscopic coherence which is the coherence between macroscopically distinct quantum states. 
We construct a measure that quantifies how global and collective the coherence of the state is. Our work also provides a first-principle way to derive well-established states like the GHZ and the NOON state as the states that maximize our measure. This new approach paves the way towards a better understanding of the Quantum-to-Classical transition. 
\end{abstract}
\maketitle

For more than a century, quantum mechanics has successfully explained a wide range of phenomena in physics. There is however one simple yet challenging question that has puzzled some of the greatest minds in physics and still remains unsolved. Namely, it is still unclear why the macroscopic world around us is classical and what the nature of the transition from the quantum physics at the microscopic level to the classical one at the macroscopic level is. This problem was manifested by Schr\"{o}dinger in the famous thought experiment of the Schr\"{o}dinger's cat \cite{schrodinger1935present}. Yet, after about a century, this problem is still the subject of active research and especially in the past two decades attracted a lot of attention\cite{leggett1980macroscopic,leggett2002testing,de2008entanglement,raeisi2011coarse,wang2013precision,farrow2015classification,frowis2018macroscopic}.

Different approaches has been taken to explain the discrepancy between the microscopic and macroscopic world. On the one hand, there are the collapse models which suggest that the theory of quantum mechanics needs to be modified to comply with our classical observations \cite{bassi2013models}. On the other hand, there are approaches that search for the solution within quantum mechanics \cite{zurek1988quantum,zurek1991decoherence,paz2002environment,zurek2003decoherence,schlosshauer2005decoherence,zurek2006decoherence,schlosshauer2007decoherence,de2012colloquium,joos2013decoherence,jeong2014coarsening}. For instance, in many cases, decoherence can explain the emergence of classical states from quantum ones. Or similarly, it has been shown that the lack of  precision could make quantum states look like  classical states \cite{raeisi2011coarse,wang2013precision,sekatski2014difficult}.

One of the key challenges of finding a resolution to the Quantum-to-Classical transition is the 
ambiguity of the problem, i.e. the lack of a clear and cohesive  picture of what  macroscopic quantum states and effects are. 

This problem has been intensively investigated for the past two decades and a variety of measures and definitions of macroscopic quantumness have been suggested \cite{leggett1980macroscopic,leggett2002testing,dur2002effective,
shimizu2002stability,bjork2004size, ukena2004appearance,morimae2005macroscopic,shimizu2005detection,cavalcanti2006signatures, cavalcanti2008criteria,marquardt2008measuring,korsbakken2010size, lee2011quantification,frowis2012measures,nimmrichter2013macroscopicity,volkoff2014measurement,sekatski2014size, oudot2015two,jeong2015macroscopic,yadin2015quantum,park2016quantum, kwon2017disturbance,  sekatski2018general}. These measures vary in approaches, formulations and applicability.
Some measures are based on comparison to well-established states such as the Greenberger-Horne-Zeilinger(GHZ) state \cite{greenberger1989going,greenberger2007going} or the Coherent Cat states \cite{schrodinger1935present,buvzek1992superpositions}.
Some other measures quantify the macroscopic quantumness of a state 
by the 
oscillations in the probability distribution with respect to some measurement. 
For example, Lee and Jeong characterized the macroscopic quantumness of photonic states based on  the intensity of oscillation frequencies of its Wigner-function \cite{jeong2015macroscopic}. 
Following this idea, Fro\"{w}is and D\"{u}r proposed 
to use Quantum Fisher Information(QFI) for characterization of macroscopic quantumness  \cite{frowis2012measures,frowis2015linking,oudot2015two}.

Lack of cohesion and diversity of  definitions and measures indicate that, although we have a better understanding of the problem, we still do not have a clear notion of what macroscopic quantumness is.

Here, we present a new approach to characterizing macroscopic quantumness. 
We start with coherence \cite{streltsov2017colloquium} which is widely believed to be the underlying feature that distinguishes quantum and classical physics \cite{streltsov2017colloquium}. 
We construct a new measure of macroscopic quantumness which is a monotone for quantum coherence that incentivize the coherence between macroscopically distinguishable states. 
This can be seen as a specific example of   the framework established by Yadin and Vedral in \cite{yadin2016general} but with the distinction that we take a first-principle approach to the problem.

Naturally, macroscopic quantum states are  expected to have relatively large amount of coherence. However, for a state to be recognized as a macroscopic quantum state, not only it should have large measurable coherence, but the coherence should also be distributed macroscopically. To clarify this, consider the following two spin states. 
\begin{align}
|\psi_{1}\rangle &= \frac{|0\rangle+|1\rangle}{\sqrt{2}}\otimes | 0\rangle^{\otimes (N-1)}\cr
|\psi_{2}\rangle &= \frac{|0\rangle ^{\otimes N}+|1\rangle^{\otimes N}}{\sqrt{2}}, 
\label{1and2}
\end{align}
where $|0\rangle$ and $|1\rangle$ correspond to up and down spins respectively. 
Most coherence measures would assign the same amount of coherence to these two states since their density matrices both have similar off-diagonal elements, both in value and number. However, the off-diagonal elements of $ |\psi_1 \rangle$ is between $|0 0 \cdots 0 \rangle$ and  $|1 0 \cdots 0 \rangle$ whereas for $|\psi_2 \rangle$  it is between $ |0 \rangle^{\otimes N}$ and  $|1 \rangle^{\otimes N}$. The difference between the two states is that, for the former, the states differ in only one spin and are not macroscopically distinguishable, whereas for the latter, they could be distinguished for large enough $N$ and with the right measurement.  
For instance, for a magnetization measurement in the z-direction, $ |0 \rangle^{\otimes N}$ gives $N(\frac{\hbar}{2})$ whereas $ |1 \rangle^{\otimes N}$ gives $-N(\frac{\hbar}{2})$.
This means that for large enough $N$, the states $| 0 \rangle^{\otimes N}$ and  $|1 \rangle^{\otimes N}$ can be distinguished with a macroscopic magnetization measurement. In this sense, it can be argued that, although both states have the same amount of coherence (quantumness), $ |\psi_2 \rangle$ has the additional property that its quantumness is distributed macroscopically, i.e. coherence is between states that are macroscopically distinguishable. 
Here we present a new characterization of macroscopic quantumness based on this notion. Namely, we start with a notion of quantumness, i.e. the coherence and add the extra requirement that it should be macroscopic. The advantage of this approach is that it does not rely on  well-established states or a phenomenological behaviour of them. Instead, to some extent, it gives a first-principle approach to the characterization of macroscopic quantumness. We will show that, this first principle approach is consistent and can characterize the well-established macroscopic quantum states properly.

We start with our notation and terminology. 
For a density matrix 
$\rho = \sum_{i,j}{\rho_{i,j}  |i\rangle \langle j|}$, the coherence is characterized by the off-diagonal elements $\rho_{i\neq j}$. We refer to $\rho_{i,j} $ as coherence elements between states $| i\rangle$ and $|j\rangle$. 

\begin{figure}
	\centering
	\includegraphics[scale=0.35]{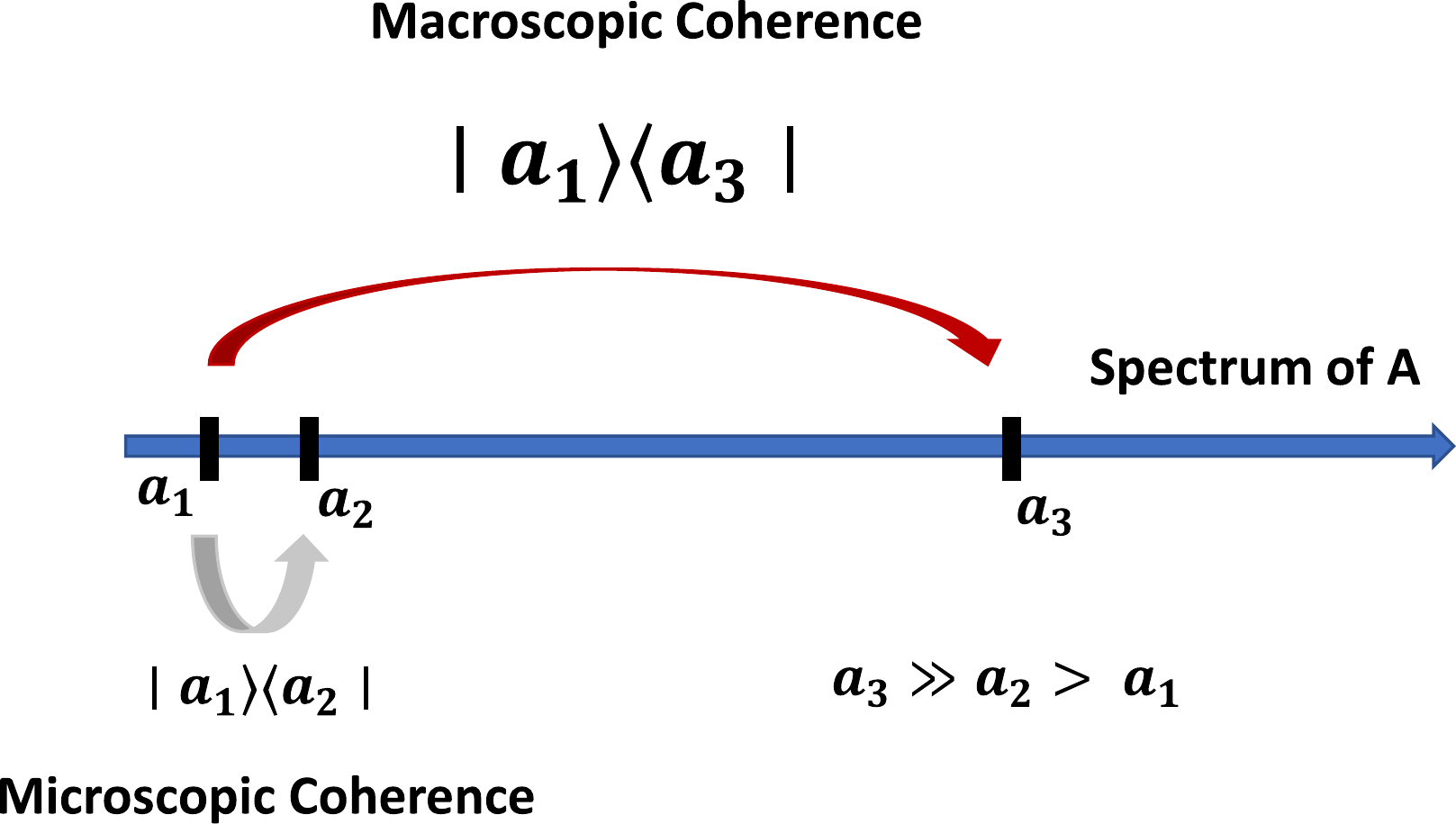}
	\caption{Schematic picture for macroscopic coherence. The spectrum 
	of an operator A, $a_1, a_2, a_3$ are depicted on the vertical axis. Two coherence elements are shown, one between the first and the second 
	eigenvalues of $A$ and one between the first and the second. The idea is that the states involved in a coherence elements may or may not be  macroscopically distinct, according to the measurement of some operator $A$. If they are macroscopically distinct, that makes the coherence macroscopic and these macroscopic coherence elements can characterize macroscopic quantumness.  }
 \label{fig:macro_coherence}
\end{figure}

Typical coherence monotones would treat all the coherence elements uniformly. However, as illustrated in the example in Eq.  \ref{1and2}, 
this approach would not be suitable for characterization of macroscopic quantum states. 
For a coherence monotones 
to captures macroscopic quantumness, it has to incentivize coherence elements between states that are more macroscopically distinct, i.e. for a coherence element $\rho_{i,j}$, the more macroscopically distinct the two states $|i\rangle$ and $|j\rangle$, the more that element should contribute to the monotone. 
To this end, we introduce ``Macroscopic Coherence'' which refers to the coherence terms  $\rho_{i,j}$ such that the states involved, i.e. $|i\rangle$ and $|j\rangle$ can be macroscopically distinguished with some measurement. 
For a schematic picture, see figure \ref{fig:macro_coherence}.

Initially, there are two ambiguity in this approach. First, it is not clear what characterizes the macroscopic distinction between the two states $|i\rangle$ and $| j\rangle$, and second, the coherence elements depend on the basis. The former is due to the unclear border between macro and micro and for this, we can rely on what is considered a macroscopic distinction in an experimental setting. The latter is because coherence is a basis-dependent quantity. 
But both of these ambiguities are expected in the characterization of macroscopic quantumness. 
For instance, for a GHZ state with $N$ spins, it is not clear for how large of a number $N$, the state would qualify as a macroscopic state. Similarly, for identifying quantumness, the basis of the measured observable is important. 
This would mean that our measure for macroscopic quantumness should depend on the  measurement. 

To quantify the macroscopic coherence,  we first need a monotone for coherence and next we need to quantify the macroscopisity of the coherence. For both of these, we need to specify the measured observable. 

Assume that the observable of interest is $A = \sum_{i}{a_i | i\rangle\langle i|}$. The eigenbasis of $A$ sets the basis for the coherence. 
For quantification of coherence we start with 
\begin{equation}\label{coherence_measure}
 \sum_{i \neq j}^{D^2-D}{|\rho_{i,j}|},   
\end{equation}
where $D$ is the dimension of the Hilbert space
\cite{streltsov2017colloquium,baumgratz2014quantifying}.

Next we need to quantify the macroscopic distinction between the states. 
Note that the elements of an orthonormal basis are mutually orthogonal and therefore, the inner product does not capture the difference between say $| 0 \rangle\langle 1| $ and $| 0 \rangle\langle N| $. 
One natural choice for the macroscopic distinction between the two states $|i\rangle$ and $| j\rangle$ is $|a_{i}-a_{j}|$, i.e. the difference between the eigenvalues associated to $| i\rangle$ and $| j\rangle$. If the difference is large enough to be resolved with a macroscopic measurement, the states $|i\rangle$ and $| j\rangle$ are macroscopically distinct. For example, for a position measurement, the states $|-1 (\mathrm{meter})\rangle$ and $| 1 (\mathrm{meter})\rangle$  would be macroscopically distinct. 
Mathematically we introduce the distance 
\begin{equation}
    d_A{(i,j)}=|a_{i}-a_{j}|.
    \label{DistanceDefinition}
\end{equation}

For a measure of macroscopic quantumness, instead of uniformly considering all of the coherence elements, we weigh them based on their corresponding distances. This penalizes contribution of coherence elements with small $d_A(i,j)$ and incentivizes the contribution from elements with large $d_A(i,j)$.

To turn the coherence monotone in Eq. \ref{coherence_measure} into a monotone for macroscopic coherence, we  add the distance to the measure which gives
\begin{equation}
\sum_{i , j}{d_A(i,j)\, |\rho_{i,j} |} . 
\label{FirstSuggestion}
\end{equation}
This incentivizes macroscopic coherence 
and suppresses the microscopic coherence.  
Note that we even included the diagonal elements that have no coherence in the sum but they are automatically suppressed by $d_A(i,i)=0$ and the sum remains unchanged.

This however has a flaw, namely, there are two ways that the measure can increase, one is by increasing the coherence (not necessarily the macroscopic elements) and the other is by increasing the macroscopicity of the coherence elements. 
For instance, consider the state 
\begin{equation}
| \psi_3 \rangle = \left(\frac{|0 \rangle+| 1\rangle}{\sqrt{2}}\right)^{\otimes N}.
\label{Si3}
\end{equation}
For large enough $N$, the quantity in Eq. \ref{FirstSuggestion} would be significantly affected by the large number of off-diagonal elements in the density matrix of $| \psi_3 \rangle$ or equivalently, large amount of coherence, although most of them are not macroscopic. 
To fix this issue, we can normalize the coherence elements. This means that instead of $|\rho_{i,j}|$, we use $\frac{|\rho_{i,j} |}{\sum_{i,j}|\rho_{i,j}|}$ which indicates the fraction of all of the elements in the density matrix corresponding to the coherence element $\rho_{i,j}$. 
This gives 
\begin{equation}
 M\left( \rho\right) =\frac{\sum_{i,j}d_A(i,j)\, |\rho_{i,j} |}{\sum_{i,j}|\rho_{i,j}|}. 
 \label{Measure}
\end{equation}
This measure can be interpreted as the average of the distance $d_A(i,j)$ over all of the different elements of density matrix. To see this more clearly, we can partition  the elements of the density matrix into classes with different values for $d_A$, i.e.
\begin{equation}
C_{\delta}= \{\rho_{i,j}|d_A(i,j)=\delta\}. 
\label{Partitioning}
\end{equation}
Based on this, we can define the following probability distribution 
\begin{equation}
P(\delta)=\frac{\sum_{\rho_{i,j}\in C_\delta}|\rho_{i,j}|}{\sum_{i,j}|\rho_{i,j}|}.
\label{P(d)Definition}
\end{equation}
This is the probability of getting a coherence element with $d_A(i,j) = \delta$. 
This probability distribution translates the measure in Eq. (\ref{Measure})  to
\begin{equation}
M(\rho) = \bar{d} = \sum_{\delta} P(\delta)\delta.
\label{MeasureAsMean}
\end{equation}
This is in fact the average distance between the states corresponding to the coherence terms $\rho_{i,j}$, i.e. $\bar{d}_A( i ,  j )$, gives a quantification for the macroscopic quantumness of the state. 

For a state with its coherence elements focused between states that are not macroscopically distant according to the observable $A$ or states with small coherence, the measure gives a small value. On the other hand, if the state has large amount of coherence and the coherence elements are mostly focused between states that can be macroscopically distinguished, the measure assigns a large amount of macroscopic quantumness to the state. 

As an example, consider the states in Eq. \ref{1and2} under the measurement of the total magnetization in the $z$ direction. 
Both states have 2 diagonal and 2 off-diagonal elements, all with the value of $1/2$. For the $\psi_1$, the distance corresponding to the off-diagonal element i.e. $d(\mid 00\cdots0\rangle,\mid 10\cdots0\rangle )$ is $1$ and this gives $M(\psi_1) = 1/2$. For the GHZ state, the distance corresponding to the off-diagonal element is $N$ which gives $M(\psi_{\mathrm{GHZ}}) = N/2$. This shows that the measure scales and grows with the system size for the GHZ state, but as expected, for $|\psi_1\rangle$, it stays constant. 
This gives a natural effective size for the system that describes the scale at which the coherence is distributed.

Here we assumed that the observable $A$ is a discrete operator, however, the measure can be extended to continuous operators by discretizing the spectrum and defining bins. 
The discretization, i.e. the bin size can be set based on the precision  of the measurements. 

This measure provides a way to define ideal states, i.e. states with maximum macroscopic coherence. 
A ``Maximum Macroscopic Quantum State(MMQS)'' can be defined as a state which maximizes the measure in Eq. (\ref{MeasureAsMean}). For instance, it is easy to show that the GHZ state is an MMQS for spin-type systems. Generally an MMQS has to be of the form 
\begin{equation}
| \psi_{\mathrm{MMQS}} \rangle = \frac{| i_{\mathrm{max}} \rangle+ e^{i\phi}| i_{\mathrm{min}}\rangle}{\sqrt{2}}, \label{MMQS}
\end{equation}
with $| i_{\mathrm{max}} \rangle$ and $| i_{\mathrm{min}} \rangle$ the states corresponding to the maximum and minimum eigenvalues of the bonded observable $A$ respectively. Note that MMQS is only well-defined when the observable $A$ is bounded. 

Apart from the phase $\phi$, the MMQS is unique if there is no degeneracy in the spectrum of $A$.  For more details, see the appendix \ref{SM-MMQS}.

This characterization, as mentioned before depends on the measured observable. 
But it is also possible to make it measurement-independent by maximizing over all possible measurements. 
However, it is often impractical and sometimes impossible to carry out the maximization \cite{frowis2018macroscopic}.  For practical purposes, it it is more convenient to specify a measurement or set of measurements and investigate the states with respect to those measurements.

This measure can also be used to define an effective size for the macroscopic quantumness of a state. This is similar to \cite{dur2002effective,korsbakken2007measurement,marquardt2008measuring,frowis2012measures,yadin2015quantum}. To this end, 
we compare the value of the measure with the corresponding MMQS. More precisely, consider a  system that is comprised of $N$ entities with state $\rho$ and assume that the measure returns a value $M(\rho)$ for the macroscopic quantumness of the state. We define the effective size $N_{eff}$ as the  size of the smallest MMQS that has the same amount of macroscopic quantumness, $M(\rho)$. Mathematically, that is
\begin{equation}
 N_{eff}(\rho) = \min\{n\mid M(\rho) \leq M(\mathrm{MMQS(n) })\},
\end{equation} 
where $\mathrm{MMQS(n) }$ is the MMQS with $n$ particles. For a spin system like the examples we considered, the $N_{eff}(\rho) = 2 M(\rho)$.

Our measure is closely connected to the work by Yadin and Vedral \cite{yadin2016general}. They presented a general framework for macroscopic quantumness in terms of coherence and put forward the idea of using a coherence measure as a tool for quantification of macroscopic quantumness. Our measure can be seen as specific example of this framework. The distinction is that instead of looking for coherence monotones that fulfil condition 4 in their work, we synthesize and construct the measure from some basic principles. Also, in our approach it is possible to replace the coherence with some other notion of quantumness if deemed necessary.

\section*{Examples}
Next we  calculate our measure for some well-known states. We consider two systems, first spin ensembles and then photonic quantum states. 

\subsection*{Spin Ensemble Systems }

We start with an ensemble of spin 1/2 particles. 
Here we consider the total magnetization which is a natural and practical measurement for spin systems. 
Without loss of generality, we take this to be  the measurement of magnetization in $z-$direction. The corresponding observable is $A = \sum_i{\sigma_z^{(i)}}$ with $\sigma_z^{(i)} = | 0 \rangle\langle 0 |- |1 \rangle\langle 1 |$ on the $i$th spin of the 
ensemble.

\begin{figure}
	\centering
	\includegraphics[scale=0.35]{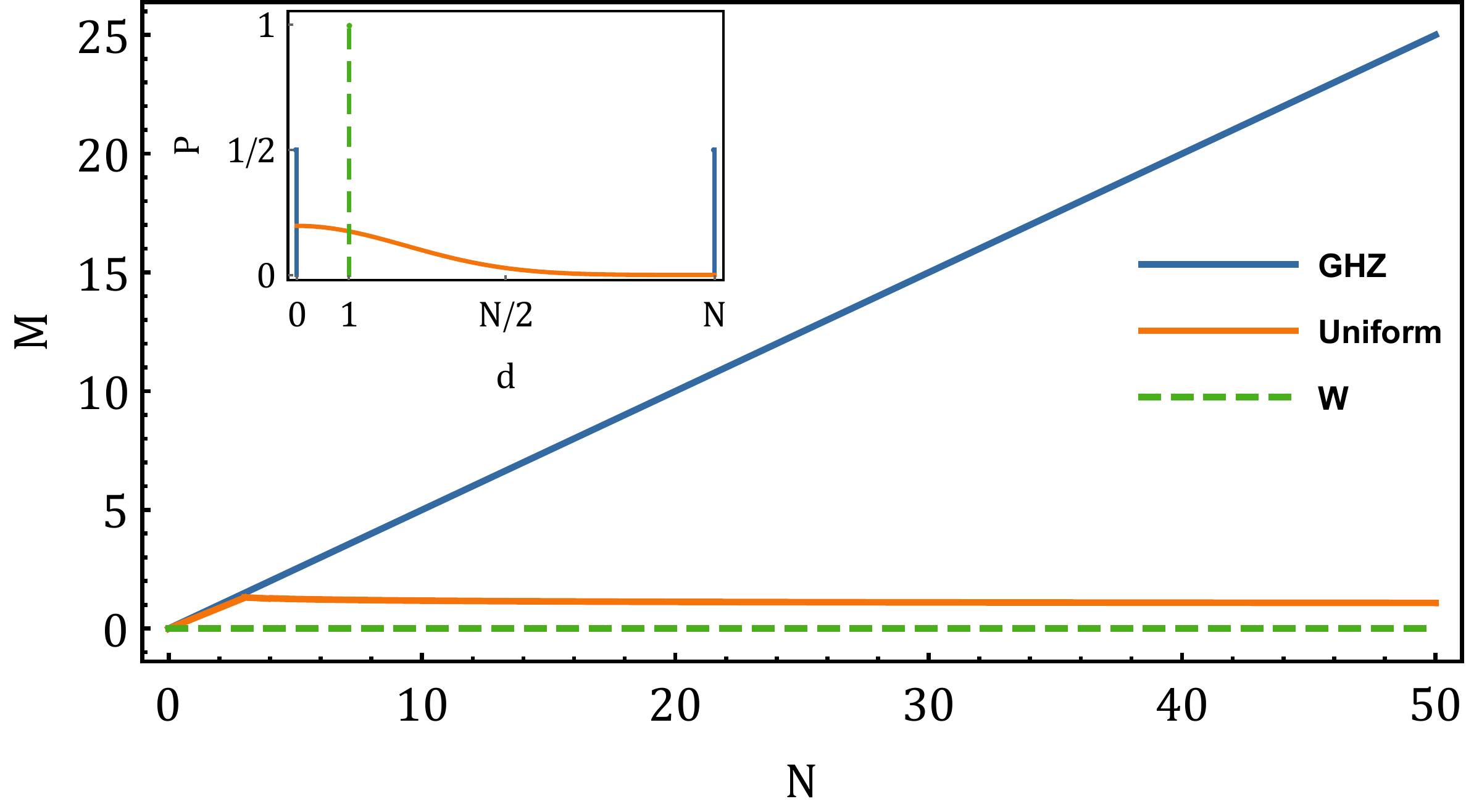}
	\caption{This plots shows how our measure for macroscopic quantumness changes with $N$ for the GHZ, Uniform and W states. $N$ is the number of spins. The inset gives a schematic plot of the probability distribution $P_{(d)}$ in Eq. \ref{P(d)Definition} for the aforementioned states. Here we are considering the total magnetization observable in $z$ direction for the measured observable. }
 \label{SpinCompare2}
\end{figure}

We start with the GHZ which is the state $|\psi_{2}\rangle$ in Eq. \ref{1and2}. As was explained before, the measure gives 
\begin{equation}
    M_{\mathrm{GHZ}}=\frac{N}{2}.  \label{MeasureGHZ}
\end{equation}
The probability distribution  $P(\delta)$ is plotted in the inset of the figure \ref{SpinCompare2} and it is clear that the mean distance is $N/2$. 

It is interesting to compare the GHZ state with $|\psi_3\rangle$. We refer to this state as the ``Uniform state''.  Similar to the GHZ state, the uniform state is macroscopic  and has non-zero coherence elements. The difference is, in contrast to the GHZ state, the coherence is not collective and each spin has its independent coherence. The probability distribution corresponding to this state is also plotted in the inset of the figure \ref{SpinCompare2}. 

For large number of spins $N$, using Stirling approximation $lnN!=NlnN-N$, the measure asymptotically  converges to
\begin{equation}
    M_{\mathrm{uni}}\approx e^{N ln (\frac{(N+\frac{1}{2})^2}{(N-1)(N+2)})}. \label{MeasureUniform}
\end{equation}
For more details, see the appendix\ref{UniformCalculation}. 

This gives
\begin{equation}
    \lim_{N\to\infty} M_{\mathrm{uni}} = 1 = O(N^{0}),  \label{LimitMeasureUniform}
\end{equation}
i.e. it converges to the constant value $1$. This is consistent with the fact the the coherence in this state is the collection of the individual coherences. 

Another interesting state is the W-state \cite{dicke1954coherence,dur2000three}. 
The W-state is given by
\begin{equation}
    |\mathrm{W} \rangle = \frac{|100...0 \rangle + |010...0\rangle +...+|000...1 \rangle}{N} \label{W}
\end{equation}

This state is an eigenstate of the magnetization in the z-direction and as a result, the distance corresponding to all of the coherence elements is zero. This means that $M_{W}=0$.

\subsection*{Photonic Systems}

Next we investigate photonic states with our measure.
For the measured observable, we consider energy or equivalently, the photon number. The state that we consider is the NOON state which is defined as
\begin{equation}
    |\mathrm{NOON}\rangle = \frac{|N\rangle|0\rangle+|0\rangle|N\rangle}{\sqrt{2}}.   \label{NOON}
\end{equation}
This state is comprised of two modes. These could be the vertical and horizontal polarization that can be separated with a polarizing beam splitter. 
The calculation of the measure is similar to the one for the GHZ state and gives 
\begin{equation}
    M_{\mathrm{NOON}}=\frac{N}{2} \label{MeasureNOON}
\end{equation}

For more examples and further details of the calculation of the measure for these examples, see the appendix. 

\begin{table}[H]
    \centering
    \begin{tabular}{|c|c|c|c|c|}
    \hline
       State  &  GHZ & NOON & Uniform & W\\
       \hline
    Measure &  $\frac{N}{2}$ & $\frac{N}{2}$ & $ 1 $ & $0$ \\
      \hline
    \end{tabular}
    \caption{The value of the measure for different states. $N$ is the number of element, i.e. spins or photons in the state.}
    \label{TableofMeasure}
\end{table}

\section{Conclusion}
 In conclusion, we presented a new approach for the characterization of macroscopic quantumness which is in fact a coherence measure. But in addition to the coherence, it also quantifies how global and collective the coherence is. Our approach can be seen as a more axiomatic alternative to established measures of macroscopic quantumness. 
 
 It also provides a first-principle approach to derive maximum macroscopic quantum states (MMQS) such as the GHZ state. We showed that the maximization of our measure over all the states would lead to MMQS. This provides a way to arrive at states such as the GHZ state in the context of macroscopic quantumnees without making any assumption about their macroscopic quantumness.

This new approach opens up a new avenue for understanding macroscopic quantumness and paves the way towards a cohesive and unified characterization of macroscopic quantumness.

 \acknowledgments{
This work is supported by the research grant system of Sharif
University of Technology (G960219).}

 \bibliographystyle{apsrev4-2}
 \bibliography{Refrences}

\begin{thebibliography}{52}%
\makeatletter
\providecommand \@ifxundefined [1]{%
 \@ifx{#1\undefined}
}%
\providecommand \@ifnum [1]{%
 \ifnum #1\expandafter \@firstoftwo
 \else \expandafter \@secondoftwo
 \fi
}%
\providecommand \@ifx [1]{%
 \ifx #1\expandafter \@firstoftwo
 \else \expandafter \@secondoftwo
 \fi
}%
\providecommand \natexlab [1]{#1}%
\providecommand \enquote  [1]{``#1''}%
\providecommand \bibnamefont  [1]{#1}%
\providecommand \bibfnamefont [1]{#1}%
\providecommand \citenamefont [1]{#1}%
\providecommand \href@noop [0]{\@secondoftwo}%
\providecommand \href [0]{\begingroup \@sanitize@url \@href}%
\providecommand \@href[1]{\@@startlink{#1}\@@href}%
\providecommand \@@href[1]{\endgroup#1\@@endlink}%
\providecommand \@sanitize@url [0]{\catcode `\\12\catcode `\$12\catcode
  `\&12\catcode `\#12\catcode `\^12\catcode `\_12\catcode `\%12\relax}%
\providecommand \@@startlink[1]{}%
\providecommand \@@endlink[0]{}%
\providecommand \url  [0]{\begingroup\@sanitize@url \@url }%
\providecommand \@url [1]{\endgroup\@href {#1}{\urlprefix }}%
\providecommand \urlprefix  [0]{URL }%
\providecommand \Eprint [0]{\href }%
\providecommand \doibase [0]{https://doi.org/}%
\providecommand \selectlanguage [0]{\@gobble}%
\providecommand \bibinfo  [0]{\@secondoftwo}%
\providecommand \bibfield  [0]{\@secondoftwo}%
\providecommand \translation [1]{[#1]}%
\providecommand \BibitemOpen [0]{}%
\providecommand \bibitemStop [0]{}%
\providecommand \bibitemNoStop [0]{.\EOS\space}%
\providecommand \EOS [0]{\spacefactor3000\relax}%
\providecommand \BibitemShut  [1]{\csname bibitem#1\endcsname}%
\let\auto@bib@innerbib\@empty
\bibitem [{\citenamefont {Schr{\"o}dinger}(1935)}]{schrodinger1935present}%
  \BibitemOpen
  \bibfield  {author} {\bibinfo {author} {\bibfnamefont {E.}~\bibnamefont
  {Schr{\"o}dinger}},\ }\href@noop {} {\bibfield  {journal} {\bibinfo
  {journal} {Die Naturwissenschaften}\ }\textbf {\bibinfo {volume} {23}},\
  \bibinfo {pages} {1} (\bibinfo {year} {1935})}\BibitemShut {NoStop}%
\bibitem [{\citenamefont {Leggett}(1980)}]{leggett1980macroscopic}%
  \BibitemOpen
  \bibfield  {author} {\bibinfo {author} {\bibfnamefont {A.~J.}\ \bibnamefont
  {Leggett}},\ }\href@noop {} {\bibfield  {journal} {\bibinfo  {journal}
  {Progress of Theoretical Physics Supplement}\ }\textbf {\bibinfo {volume}
  {69}},\ \bibinfo {pages} {80} (\bibinfo {year} {1980})}\BibitemShut {NoStop}%
\bibitem [{\citenamefont {Leggett}(2002)}]{leggett2002testing}%
  \BibitemOpen
  \bibfield  {author} {\bibinfo {author} {\bibfnamefont {A.~J.}\ \bibnamefont
  {Leggett}},\ }\href@noop {} {\bibfield  {journal} {\bibinfo  {journal}
  {Journal of Physics: Condensed Matter}\ }\textbf {\bibinfo {volume} {14}},\
  \bibinfo {pages} {R415} (\bibinfo {year} {2002})}\BibitemShut {NoStop}%
\bibitem [{\citenamefont {De~Martini}\ \emph {et~al.}(2008)\citenamefont
  {De~Martini}, \citenamefont {Sciarrino},\ and\ \citenamefont
  {Vitelli}}]{de2008entanglement}%
  \BibitemOpen
  \bibfield  {author} {\bibinfo {author} {\bibfnamefont {F.}~\bibnamefont
  {De~Martini}}, \bibinfo {author} {\bibfnamefont {F.}~\bibnamefont
  {Sciarrino}},\ and\ \bibinfo {author} {\bibfnamefont {C.}~\bibnamefont
  {Vitelli}},\ }\href@noop {} {\bibfield  {journal} {\bibinfo  {journal}
  {Physical Review Letters}\ }\textbf {\bibinfo {volume} {100}},\ \bibinfo
  {pages} {253601} (\bibinfo {year} {2008})}\BibitemShut {NoStop}%
\bibitem [{\citenamefont {Raeisi}\ \emph {et~al.}(2011)\citenamefont {Raeisi},
  \citenamefont {Sekatski},\ and\ \citenamefont {Simon}}]{raeisi2011coarse}%
  \BibitemOpen
  \bibfield  {author} {\bibinfo {author} {\bibfnamefont {S.}~\bibnamefont
  {Raeisi}}, \bibinfo {author} {\bibfnamefont {P.}~\bibnamefont {Sekatski}},\
  and\ \bibinfo {author} {\bibfnamefont {C.}~\bibnamefont {Simon}},\
  }\href@noop {} {\bibfield  {journal} {\bibinfo  {journal} {Physical Review
  Letters}\ }\textbf {\bibinfo {volume} {107}},\ \bibinfo {pages} {250401}
  (\bibinfo {year} {2011})}\BibitemShut {NoStop}%
\bibitem [{\citenamefont {Wang}\ \emph {et~al.}(2013)\citenamefont {Wang},
  \citenamefont {Ghobadi}, \citenamefont {Raeisi},\ and\ \citenamefont
  {Simon}}]{wang2013precision}%
  \BibitemOpen
  \bibfield  {author} {\bibinfo {author} {\bibfnamefont {T.}~\bibnamefont
  {Wang}}, \bibinfo {author} {\bibfnamefont {R.}~\bibnamefont {Ghobadi}},
  \bibinfo {author} {\bibfnamefont {S.}~\bibnamefont {Raeisi}},\ and\ \bibinfo
  {author} {\bibfnamefont {C.}~\bibnamefont {Simon}},\ }\href@noop {}
  {\bibfield  {journal} {\bibinfo  {journal} {Physical Review A}\ }\textbf
  {\bibinfo {volume} {88}},\ \bibinfo {pages} {062114} (\bibinfo {year}
  {2013})}\BibitemShut {NoStop}%
\bibitem [{\citenamefont {Farrow}\ and\ \citenamefont
  {Vedral}(2015)}]{farrow2015classification}%
  \BibitemOpen
  \bibfield  {author} {\bibinfo {author} {\bibfnamefont {T.}~\bibnamefont
  {Farrow}}\ and\ \bibinfo {author} {\bibfnamefont {V.}~\bibnamefont
  {Vedral}},\ }\href@noop {} {\bibfield  {journal} {\bibinfo  {journal} {Optics
  Communications}\ }\textbf {\bibinfo {volume} {337}},\ \bibinfo {pages} {22}
  (\bibinfo {year} {2015})}\BibitemShut {NoStop}%
\bibitem [{\citenamefont {Fr{\"o}wis}\ \emph {et~al.}(2018)\citenamefont
  {Fr{\"o}wis}, \citenamefont {Sekatski}, \citenamefont {D{\"u}r},
  \citenamefont {Gisin},\ and\ \citenamefont
  {Sangouard}}]{frowis2018macroscopic}%
  \BibitemOpen
  \bibfield  {author} {\bibinfo {author} {\bibfnamefont {F.}~\bibnamefont
  {Fr{\"o}wis}}, \bibinfo {author} {\bibfnamefont {P.}~\bibnamefont
  {Sekatski}}, \bibinfo {author} {\bibfnamefont {W.}~\bibnamefont {D{\"u}r}},
  \bibinfo {author} {\bibfnamefont {N.}~\bibnamefont {Gisin}},\ and\ \bibinfo
  {author} {\bibfnamefont {N.}~\bibnamefont {Sangouard}},\ }\href@noop {}
  {\bibfield  {journal} {\bibinfo  {journal} {Reviews of Modern Physics}\
  }\textbf {\bibinfo {volume} {90}},\ \bibinfo {pages} {025004} (\bibinfo
  {year} {2018})}\BibitemShut {NoStop}%
\bibitem [{\citenamefont {Bassi}\ \emph {et~al.}(2013)\citenamefont {Bassi},
  \citenamefont {Lochan}, \citenamefont {Satin}, \citenamefont {Singh},\ and\
  \citenamefont {Ulbricht}}]{bassi2013models}%
  \BibitemOpen
  \bibfield  {author} {\bibinfo {author} {\bibfnamefont {A.}~\bibnamefont
  {Bassi}}, \bibinfo {author} {\bibfnamefont {K.}~\bibnamefont {Lochan}},
  \bibinfo {author} {\bibfnamefont {S.}~\bibnamefont {Satin}}, \bibinfo
  {author} {\bibfnamefont {T.~P.}\ \bibnamefont {Singh}},\ and\ \bibinfo
  {author} {\bibfnamefont {H.}~\bibnamefont {Ulbricht}},\ }\href@noop {}
  {\bibfield  {journal} {\bibinfo  {journal} {Reviews of Modern Physics}\
  }\textbf {\bibinfo {volume} {85}},\ \bibinfo {pages} {471} (\bibinfo {year}
  {2013})}\BibitemShut {NoStop}%
\bibitem [{\citenamefont {Zurek}(1988)}]{zurek1988quantum}%
  \BibitemOpen
  \bibfield  {author} {\bibinfo {author} {\bibfnamefont {W.~H.}\ \bibnamefont
  {Zurek}},\ }\href@noop {} {\emph {\bibinfo {title} {Quantum Measurements and
  the Environment Induced Transition from Quantum to Classical}}},\ \bibinfo
  {type} {Tech. Rep.}\ (\bibinfo  {institution} {Los Alamos National Lab., NM
  (USA)},\ \bibinfo {year} {1988})\BibitemShut {NoStop}%
\bibitem [{\citenamefont {Zurek}(1991)}]{zurek1991decoherence}%
  \BibitemOpen
  \bibfield  {author} {\bibinfo {author} {\bibfnamefont {W.}~\bibnamefont
  {Zurek}},\ }\href@noop {} {\bibinfo {title} {Decoherence and the transition
  from quantum to classical physics today}} (\bibinfo {year}
  {1991})\BibitemShut {NoStop}%
\bibitem [{\citenamefont {Paz}\ and\ \citenamefont
  {Zurek}(2002)}]{paz2002environment}%
  \BibitemOpen
  \bibfield  {author} {\bibinfo {author} {\bibfnamefont {J.~P.}\ \bibnamefont
  {Paz}}\ and\ \bibinfo {author} {\bibfnamefont {W.~H.}\ \bibnamefont
  {Zurek}},\ }in\ \href@noop {} {\emph {\bibinfo {booktitle} {Fundamentals of
  Quantum Information}}}\ (\bibinfo  {publisher} {Springer},\ \bibinfo {year}
  {2002})\ pp.\ \bibinfo {pages} {77--148}\BibitemShut {NoStop}%
\bibitem [{\citenamefont {Zurek}(2003)}]{zurek2003decoherence}%
  \BibitemOpen
  \bibfield  {author} {\bibinfo {author} {\bibfnamefont {W.~H.}\ \bibnamefont
  {Zurek}},\ }\href@noop {} {\bibfield  {journal} {\bibinfo  {journal} {Reviews
  of Modern Physics}\ }\textbf {\bibinfo {volume} {75}},\ \bibinfo {pages}
  {715} (\bibinfo {year} {2003})}\BibitemShut {NoStop}%
\bibitem [{\citenamefont {Schlosshauer}(2005)}]{schlosshauer2005decoherence}%
  \BibitemOpen
  \bibfield  {author} {\bibinfo {author} {\bibfnamefont {M.}~\bibnamefont
  {Schlosshauer}},\ }\href@noop {} {\bibfield  {journal} {\bibinfo  {journal}
  {Reviews of Modern physics}\ }\textbf {\bibinfo {volume} {76}},\ \bibinfo
  {pages} {1267} (\bibinfo {year} {2005})}\BibitemShut {NoStop}%
\bibitem [{\citenamefont {Zurek}(2006)}]{zurek2006decoherence}%
  \BibitemOpen
  \bibfield  {author} {\bibinfo {author} {\bibfnamefont {W.~H.}\ \bibnamefont
  {Zurek}},\ }in\ \href@noop {} {\emph {\bibinfo {booktitle} {Quantum
  Decoherence}}}\ (\bibinfo  {publisher} {Springer},\ \bibinfo {year} {2006})\
  pp.\ \bibinfo {pages} {1--31}\BibitemShut {NoStop}%
\bibitem [{\citenamefont {Schlosshauer}(2007)}]{schlosshauer2007decoherence}%
  \BibitemOpen
  \bibfield  {author} {\bibinfo {author} {\bibfnamefont {M.~A.}\ \bibnamefont
  {Schlosshauer}},\ }\href@noop {} {\emph {\bibinfo {title} {Decoherence: and
  the quantum-to-classical transition}}}\ (\bibinfo  {publisher} {Springer
  Science \& Business Media},\ \bibinfo {year} {2007})\BibitemShut {NoStop}%
\bibitem [{\citenamefont {De~Martini}\ and\ \citenamefont
  {Sciarrino}(2012)}]{de2012colloquium}%
  \BibitemOpen
  \bibfield  {author} {\bibinfo {author} {\bibfnamefont {F.}~\bibnamefont
  {De~Martini}}\ and\ \bibinfo {author} {\bibfnamefont {F.}~\bibnamefont
  {Sciarrino}},\ }\href@noop {} {\bibfield  {journal} {\bibinfo  {journal}
  {Reviews of Modern Physics}\ }\textbf {\bibinfo {volume} {84}},\ \bibinfo
  {pages} {1765} (\bibinfo {year} {2012})}\BibitemShut {NoStop}%
\bibitem [{\citenamefont {Joos}\ \emph {et~al.}(2013)\citenamefont {Joos},
  \citenamefont {Zeh}, \citenamefont {Kiefer}, \citenamefont {Giulini},
  \citenamefont {Kupsch},\ and\ \citenamefont
  {Stamatescu}}]{joos2013decoherence}%
  \BibitemOpen
  \bibfield  {author} {\bibinfo {author} {\bibfnamefont {E.}~\bibnamefont
  {Joos}}, \bibinfo {author} {\bibfnamefont {H.~D.}\ \bibnamefont {Zeh}},
  \bibinfo {author} {\bibfnamefont {C.}~\bibnamefont {Kiefer}}, \bibinfo
  {author} {\bibfnamefont {D.~J.}\ \bibnamefont {Giulini}}, \bibinfo {author}
  {\bibfnamefont {J.}~\bibnamefont {Kupsch}},\ and\ \bibinfo {author}
  {\bibfnamefont {I.-O.}\ \bibnamefont {Stamatescu}},\ }\href@noop {} {\emph
  {\bibinfo {title} {Decoherence and the appearance of a classical world in
  quantum theory}}}\ (\bibinfo  {publisher} {Springer Science \& Business
  Media},\ \bibinfo {year} {2013})\BibitemShut {NoStop}%
\bibitem [{\citenamefont {Jeong}\ \emph {et~al.}(2014)\citenamefont {Jeong},
  \citenamefont {Lim},\ and\ \citenamefont {Kim}}]{jeong2014coarsening}%
  \BibitemOpen
  \bibfield  {author} {\bibinfo {author} {\bibfnamefont {H.}~\bibnamefont
  {Jeong}}, \bibinfo {author} {\bibfnamefont {Y.}~\bibnamefont {Lim}},\ and\
  \bibinfo {author} {\bibfnamefont {M.}~\bibnamefont {Kim}},\ }\href@noop {}
  {\bibfield  {journal} {\bibinfo  {journal} {Physical Review Letters}\
  }\textbf {\bibinfo {volume} {112}},\ \bibinfo {pages} {010402} (\bibinfo
  {year} {2014})}\BibitemShut {NoStop}%
\bibitem [{\citenamefont {Sekatski}\ \emph
  {et~al.}(2014{\natexlab{a}})\citenamefont {Sekatski}, \citenamefont {Gisin},\
  and\ \citenamefont {Sangouard}}]{sekatski2014difficult}%
  \BibitemOpen
  \bibfield  {author} {\bibinfo {author} {\bibfnamefont {P.}~\bibnamefont
  {Sekatski}}, \bibinfo {author} {\bibfnamefont {N.}~\bibnamefont {Gisin}},\
  and\ \bibinfo {author} {\bibfnamefont {N.}~\bibnamefont {Sangouard}},\
  }\href@noop {} {\bibfield  {journal} {\bibinfo  {journal} {Physical Review
  Letters}\ }\textbf {\bibinfo {volume} {113}},\ \bibinfo {pages} {090403}
  (\bibinfo {year} {2014}{\natexlab{a}})}\BibitemShut {NoStop}%
\bibitem [{\citenamefont {D{\"u}r}\ \emph {et~al.}(2002)\citenamefont
  {D{\"u}r}, \citenamefont {Simon},\ and\ \citenamefont
  {Cirac}}]{dur2002effective}%
  \BibitemOpen
  \bibfield  {author} {\bibinfo {author} {\bibfnamefont {W.}~\bibnamefont
  {D{\"u}r}}, \bibinfo {author} {\bibfnamefont {C.}~\bibnamefont {Simon}},\
  and\ \bibinfo {author} {\bibfnamefont {J.~I.}\ \bibnamefont {Cirac}},\
  }\href@noop {} {\bibfield  {journal} {\bibinfo  {journal} {Physical Review
  Letters}\ }\textbf {\bibinfo {volume} {89}},\ \bibinfo {pages} {210402}
  (\bibinfo {year} {2002})}\BibitemShut {NoStop}%
\bibitem [{\citenamefont {Shimizu}\ and\ \citenamefont
  {Miyadera}(2002)}]{shimizu2002stability}%
  \BibitemOpen
  \bibfield  {author} {\bibinfo {author} {\bibfnamefont {A.}~\bibnamefont
  {Shimizu}}\ and\ \bibinfo {author} {\bibfnamefont {T.}~\bibnamefont
  {Miyadera}},\ }\href@noop {} {\bibfield  {journal} {\bibinfo  {journal}
  {Physical Review Letters}\ }\textbf {\bibinfo {volume} {89}},\ \bibinfo
  {pages} {270403} (\bibinfo {year} {2002})}\BibitemShut {NoStop}%
\bibitem [{\citenamefont {Bj{\"o}rk}\ and\ \citenamefont
  {Mana}(2004)}]{bjork2004size}%
  \BibitemOpen
  \bibfield  {author} {\bibinfo {author} {\bibfnamefont {G.}~\bibnamefont
  {Bj{\"o}rk}}\ and\ \bibinfo {author} {\bibfnamefont {P.~G.~L.}\ \bibnamefont
  {Mana}},\ }\href@noop {} {\bibfield  {journal} {\bibinfo  {journal} {Journal
  of Optics B: Quantum and Semiclassical Optics}\ }\textbf {\bibinfo {volume}
  {6}},\ \bibinfo {pages} {429} (\bibinfo {year} {2004})}\BibitemShut {NoStop}%
\bibitem [{\citenamefont {Ukena}\ and\ \citenamefont
  {Shimizu}(2004)}]{ukena2004appearance}%
  \BibitemOpen
  \bibfield  {author} {\bibinfo {author} {\bibfnamefont {A.}~\bibnamefont
  {Ukena}}\ and\ \bibinfo {author} {\bibfnamefont {A.}~\bibnamefont
  {Shimizu}},\ }\href@noop {} {\bibfield  {journal} {\bibinfo  {journal}
  {Physical Review A}\ }\textbf {\bibinfo {volume} {69}},\ \bibinfo {pages}
  {022301} (\bibinfo {year} {2004})}\BibitemShut {NoStop}%
\bibitem [{\citenamefont {Morimae}\ \emph {et~al.}(2005)\citenamefont
  {Morimae}, \citenamefont {Sugita},\ and\ \citenamefont
  {Shimizu}}]{morimae2005macroscopic}%
  \BibitemOpen
  \bibfield  {author} {\bibinfo {author} {\bibfnamefont {T.}~\bibnamefont
  {Morimae}}, \bibinfo {author} {\bibfnamefont {A.}~\bibnamefont {Sugita}},\
  and\ \bibinfo {author} {\bibfnamefont {A.}~\bibnamefont {Shimizu}},\
  }\href@noop {} {\bibfield  {journal} {\bibinfo  {journal} {Physical Review
  A}\ }\textbf {\bibinfo {volume} {71}},\ \bibinfo {pages} {032317} (\bibinfo
  {year} {2005})}\BibitemShut {NoStop}%
\bibitem [{\citenamefont {Shimizu}\ and\ \citenamefont
  {Morimae}(2005)}]{shimizu2005detection}%
  \BibitemOpen
  \bibfield  {author} {\bibinfo {author} {\bibfnamefont {A.}~\bibnamefont
  {Shimizu}}\ and\ \bibinfo {author} {\bibfnamefont {T.}~\bibnamefont
  {Morimae}},\ }\href@noop {} {\bibfield  {journal} {\bibinfo  {journal}
  {Physical Review Letters}\ }\textbf {\bibinfo {volume} {95}},\ \bibinfo
  {pages} {090401} (\bibinfo {year} {2005})}\BibitemShut {NoStop}%
\bibitem [{\citenamefont {Cavalcanti}\ and\ \citenamefont
  {Reid}(2006)}]{cavalcanti2006signatures}%
  \BibitemOpen
  \bibfield  {author} {\bibinfo {author} {\bibfnamefont {E.}~\bibnamefont
  {Cavalcanti}}\ and\ \bibinfo {author} {\bibfnamefont {M.}~\bibnamefont
  {Reid}},\ }\href@noop {} {\bibfield  {journal} {\bibinfo  {journal} {Physical
  Review Letters}\ }\textbf {\bibinfo {volume} {97}},\ \bibinfo {pages}
  {170405} (\bibinfo {year} {2006})}\BibitemShut {NoStop}%
\bibitem [{\citenamefont {Cavalcanti}\ and\ \citenamefont
  {Reid}(2008)}]{cavalcanti2008criteria}%
  \BibitemOpen
  \bibfield  {author} {\bibinfo {author} {\bibfnamefont {E.}~\bibnamefont
  {Cavalcanti}}\ and\ \bibinfo {author} {\bibfnamefont {M.}~\bibnamefont
  {Reid}},\ }\href@noop {} {\bibfield  {journal} {\bibinfo  {journal} {Physical
  Review A}\ }\textbf {\bibinfo {volume} {77}},\ \bibinfo {pages} {062108}
  (\bibinfo {year} {2008})}\BibitemShut {NoStop}%
\bibitem [{\citenamefont {Marquardt}\ \emph {et~al.}(2008)\citenamefont
  {Marquardt}, \citenamefont {Abel},\ and\ \citenamefont {von
  Delft}}]{marquardt2008measuring}%
  \BibitemOpen
  \bibfield  {author} {\bibinfo {author} {\bibfnamefont {F.}~\bibnamefont
  {Marquardt}}, \bibinfo {author} {\bibfnamefont {B.}~\bibnamefont {Abel}},\
  and\ \bibinfo {author} {\bibfnamefont {J.}~\bibnamefont {von Delft}},\
  }\href@noop {} {\bibfield  {journal} {\bibinfo  {journal} {Physical Review
  A}\ }\textbf {\bibinfo {volume} {78}},\ \bibinfo {pages} {012109} (\bibinfo
  {year} {2008})}\BibitemShut {NoStop}%
\bibitem [{\citenamefont {Korsbakken}\ \emph {et~al.}(2010)\citenamefont
  {Korsbakken}, \citenamefont {Wilhelm},\ and\ \citenamefont
  {Whaley}}]{korsbakken2010size}%
  \BibitemOpen
  \bibfield  {author} {\bibinfo {author} {\bibfnamefont {J.}~\bibnamefont
  {Korsbakken}}, \bibinfo {author} {\bibfnamefont {F.}~\bibnamefont
  {Wilhelm}},\ and\ \bibinfo {author} {\bibfnamefont {K.}~\bibnamefont
  {Whaley}},\ }\href@noop {} {\bibfield  {journal} {\bibinfo  {journal} {EPL
  (Europhysics Letters)}\ }\textbf {\bibinfo {volume} {89}},\ \bibinfo {pages}
  {30003} (\bibinfo {year} {2010})}\BibitemShut {NoStop}%
\bibitem [{\citenamefont {Lee}\ and\ \citenamefont
  {Jeong}(2011)}]{lee2011quantification}%
  \BibitemOpen
  \bibfield  {author} {\bibinfo {author} {\bibfnamefont {C.-W.}\ \bibnamefont
  {Lee}}\ and\ \bibinfo {author} {\bibfnamefont {H.}~\bibnamefont {Jeong}},\
  }\href@noop {} {\bibfield  {journal} {\bibinfo  {journal} {Physical Review
  Letters}\ }\textbf {\bibinfo {volume} {106}},\ \bibinfo {pages} {220401}
  (\bibinfo {year} {2011})}\BibitemShut {NoStop}%
\bibitem [{\citenamefont {Fr{\"o}wis}\ and\ \citenamefont
  {D{\"u}r}(2012)}]{frowis2012measures}%
  \BibitemOpen
  \bibfield  {author} {\bibinfo {author} {\bibfnamefont {F.}~\bibnamefont
  {Fr{\"o}wis}}\ and\ \bibinfo {author} {\bibfnamefont {W.}~\bibnamefont
  {D{\"u}r}},\ }\href@noop {} {\bibfield  {journal} {\bibinfo  {journal} {New
  Journal of Physics}\ }\textbf {\bibinfo {volume} {14}},\ \bibinfo {pages}
  {093039} (\bibinfo {year} {2012})}\BibitemShut {NoStop}%
\bibitem [{\citenamefont {Nimmrichter}\ and\ \citenamefont
  {Hornberger}(2013)}]{nimmrichter2013macroscopicity}%
  \BibitemOpen
  \bibfield  {author} {\bibinfo {author} {\bibfnamefont {S.}~\bibnamefont
  {Nimmrichter}}\ and\ \bibinfo {author} {\bibfnamefont {K.}~\bibnamefont
  {Hornberger}},\ }\href@noop {} {\bibfield  {journal} {\bibinfo  {journal}
  {Physical Review Letters}\ }\textbf {\bibinfo {volume} {110}},\ \bibinfo
  {pages} {160403} (\bibinfo {year} {2013})}\BibitemShut {NoStop}%
\bibitem [{\citenamefont {Volkoff}\ and\ \citenamefont
  {Whaley}(2014)}]{volkoff2014measurement}%
  \BibitemOpen
  \bibfield  {author} {\bibinfo {author} {\bibfnamefont {T.}~\bibnamefont
  {Volkoff}}\ and\ \bibinfo {author} {\bibfnamefont {K.}~\bibnamefont
  {Whaley}},\ }\href@noop {} {\bibfield  {journal} {\bibinfo  {journal}
  {Physical Review A}\ }\textbf {\bibinfo {volume} {89}},\ \bibinfo {pages}
  {012122} (\bibinfo {year} {2014})}\BibitemShut {NoStop}%
\bibitem [{\citenamefont {Sekatski}\ \emph
  {et~al.}(2014{\natexlab{b}})\citenamefont {Sekatski}, \citenamefont
  {Sangouard},\ and\ \citenamefont {Gisin}}]{sekatski2014size}%
  \BibitemOpen
  \bibfield  {author} {\bibinfo {author} {\bibfnamefont {P.}~\bibnamefont
  {Sekatski}}, \bibinfo {author} {\bibfnamefont {N.}~\bibnamefont
  {Sangouard}},\ and\ \bibinfo {author} {\bibfnamefont {N.}~\bibnamefont
  {Gisin}},\ }\href@noop {} {\bibfield  {journal} {\bibinfo  {journal}
  {Physical Review A}\ }\textbf {\bibinfo {volume} {89}},\ \bibinfo {pages}
  {012116} (\bibinfo {year} {2014}{\natexlab{b}})}\BibitemShut {NoStop}%
\bibitem [{\citenamefont {Oudot}\ \emph {et~al.}(2015)\citenamefont {Oudot},
  \citenamefont {Sekatski}, \citenamefont {Fr{\"o}wis}, \citenamefont {Gisin},\
  and\ \citenamefont {Sangouard}}]{oudot2015two}%
  \BibitemOpen
  \bibfield  {author} {\bibinfo {author} {\bibfnamefont {E.}~\bibnamefont
  {Oudot}}, \bibinfo {author} {\bibfnamefont {P.}~\bibnamefont {Sekatski}},
  \bibinfo {author} {\bibfnamefont {F.}~\bibnamefont {Fr{\"o}wis}}, \bibinfo
  {author} {\bibfnamefont {N.}~\bibnamefont {Gisin}},\ and\ \bibinfo {author}
  {\bibfnamefont {N.}~\bibnamefont {Sangouard}},\ }\href@noop {} {\bibfield
  {journal} {\bibinfo  {journal} {JOSA B}\ }\textbf {\bibinfo {volume} {32}},\
  \bibinfo {pages} {2190} (\bibinfo {year} {2015})}\BibitemShut {NoStop}%
\bibitem [{\citenamefont {Jeong}\ and\ \citenamefont
  {Sasaki}(2015)}]{jeong2015macroscopic}%
  \BibitemOpen
  \bibfield  {author} {\bibinfo {author} {\bibfnamefont {H.}~\bibnamefont
  {Jeong}}\ and\ \bibinfo {author} {\bibfnamefont {M.}~\bibnamefont {Sasaki}},\
  }\href@noop {} {\bibfield  {journal} {\bibinfo  {journal} {Optics
  Communications}\ }\textbf {\bibinfo {volume} {337}},\ \bibinfo {pages} {1}
  (\bibinfo {year} {2015})}\BibitemShut {NoStop}%
\bibitem [{\citenamefont {Yadin}\ and\ \citenamefont
  {Vedral}(2015)}]{yadin2015quantum}%
  \BibitemOpen
  \bibfield  {author} {\bibinfo {author} {\bibfnamefont {B.}~\bibnamefont
  {Yadin}}\ and\ \bibinfo {author} {\bibfnamefont {V.}~\bibnamefont {Vedral}},\
  }\href@noop {} {\bibfield  {journal} {\bibinfo  {journal} {Physical Review
  A}\ }\textbf {\bibinfo {volume} {92}},\ \bibinfo {pages} {022356} (\bibinfo
  {year} {2015})}\BibitemShut {NoStop}%
\bibitem [{\citenamefont {Park}\ \emph {et~al.}(2016)\citenamefont {Park},
  \citenamefont {Kang}, \citenamefont {Lee}, \citenamefont {Bang},
  \citenamefont {Lee},\ and\ \citenamefont {Jeong}}]{park2016quantum}%
  \BibitemOpen
  \bibfield  {author} {\bibinfo {author} {\bibfnamefont {C.-Y.}\ \bibnamefont
  {Park}}, \bibinfo {author} {\bibfnamefont {M.}~\bibnamefont {Kang}}, \bibinfo
  {author} {\bibfnamefont {C.-W.}\ \bibnamefont {Lee}}, \bibinfo {author}
  {\bibfnamefont {J.}~\bibnamefont {Bang}}, \bibinfo {author} {\bibfnamefont
  {S.-W.}\ \bibnamefont {Lee}},\ and\ \bibinfo {author} {\bibfnamefont
  {H.}~\bibnamefont {Jeong}},\ }\href@noop {} {\bibfield  {journal} {\bibinfo
  {journal} {Physical Review A}\ }\textbf {\bibinfo {volume} {94}},\ \bibinfo
  {pages} {052105} (\bibinfo {year} {2016})}\BibitemShut {NoStop}%
\bibitem [{\citenamefont {Kwon}\ \emph {et~al.}(2017)\citenamefont {Kwon},
  \citenamefont {Park}, \citenamefont {Tan},\ and\ \citenamefont
  {Jeong}}]{kwon2017disturbance}%
  \BibitemOpen
  \bibfield  {author} {\bibinfo {author} {\bibfnamefont {H.}~\bibnamefont
  {Kwon}}, \bibinfo {author} {\bibfnamefont {C.-Y.}\ \bibnamefont {Park}},
  \bibinfo {author} {\bibfnamefont {K.~C.}\ \bibnamefont {Tan}},\ and\ \bibinfo
  {author} {\bibfnamefont {H.}~\bibnamefont {Jeong}},\ }\href@noop {}
  {\bibfield  {journal} {\bibinfo  {journal} {New Journal of Physics}\ }\textbf
  {\bibinfo {volume} {19}},\ \bibinfo {pages} {043024} (\bibinfo {year}
  {2017})}\BibitemShut {NoStop}%
\bibitem [{\citenamefont {Sekatski}\ \emph {et~al.}(2018)\citenamefont
  {Sekatski}, \citenamefont {Yadin}, \citenamefont {Renou}, \citenamefont
  {D{\"u}r}, \citenamefont {Gisin},\ and\ \citenamefont
  {Fr{\"o}wis}}]{sekatski2018general}%
  \BibitemOpen
  \bibfield  {author} {\bibinfo {author} {\bibfnamefont {P.}~\bibnamefont
  {Sekatski}}, \bibinfo {author} {\bibfnamefont {B.}~\bibnamefont {Yadin}},
  \bibinfo {author} {\bibfnamefont {M.-O.}\ \bibnamefont {Renou}}, \bibinfo
  {author} {\bibfnamefont {W.}~\bibnamefont {D{\"u}r}}, \bibinfo {author}
  {\bibfnamefont {N.}~\bibnamefont {Gisin}},\ and\ \bibinfo {author}
  {\bibfnamefont {F.}~\bibnamefont {Fr{\"o}wis}},\ }\href@noop {} {\bibfield
  {journal} {\bibinfo  {journal} {New Journal of Physics}\ }\textbf {\bibinfo
  {volume} {20}},\ \bibinfo {pages} {013025} (\bibinfo {year}
  {2018})}\BibitemShut {NoStop}%
\bibitem [{\citenamefont {Greenberger}\ \emph {et~al.}(1989)\citenamefont
  {Greenberger}, \citenamefont {Horne},\ and\ \citenamefont
  {Zeilinger}}]{greenberger1989going}%
  \BibitemOpen
  \bibfield  {author} {\bibinfo {author} {\bibfnamefont {D.~M.}\ \bibnamefont
  {Greenberger}}, \bibinfo {author} {\bibfnamefont {M.~A.}\ \bibnamefont
  {Horne}},\ and\ \bibinfo {author} {\bibfnamefont {A.}~\bibnamefont
  {Zeilinger}},\ }in\ \href@noop {} {\emph {\bibinfo {booktitle} {Bell’s
  theorem, quantum theory and conceptions of the universe}}}\ (\bibinfo
  {publisher} {Springer},\ \bibinfo {year} {1989})\ pp.\ \bibinfo {pages}
  {69--72}\BibitemShut {NoStop}%
\bibitem [{\citenamefont {Greenberger}\ \emph {et~al.}(2007)\citenamefont
  {Greenberger}, \citenamefont {Horne},\ and\ \citenamefont
  {Zeilinger}}]{greenberger2007going}%
  \BibitemOpen
  \bibfield  {author} {\bibinfo {author} {\bibfnamefont {D.~M.}\ \bibnamefont
  {Greenberger}}, \bibinfo {author} {\bibfnamefont {M.~A.}\ \bibnamefont
  {Horne}},\ and\ \bibinfo {author} {\bibfnamefont {A.}~\bibnamefont
  {Zeilinger}},\ }\href@noop {} {\bibfield  {journal} {\bibinfo  {journal}
  {arXiv preprint arXiv:0712.0921}\ } (\bibinfo {year} {2007})}\BibitemShut
  {NoStop}%
\bibitem [{\citenamefont {Bu{\v{z}}ek}\ \emph {et~al.}(1992)\citenamefont
  {Bu{\v{z}}ek}, \citenamefont {Vidiella-Barranco},\ and\ \citenamefont
  {Knight}}]{buvzek1992superpositions}%
  \BibitemOpen
  \bibfield  {author} {\bibinfo {author} {\bibfnamefont {V.}~\bibnamefont
  {Bu{\v{z}}ek}}, \bibinfo {author} {\bibfnamefont {A.}~\bibnamefont
  {Vidiella-Barranco}},\ and\ \bibinfo {author} {\bibfnamefont {P.~L.}\
  \bibnamefont {Knight}},\ }\href@noop {} {\bibfield  {journal} {\bibinfo
  {journal} {Physical Review A}\ }\textbf {\bibinfo {volume} {45}},\ \bibinfo
  {pages} {6570} (\bibinfo {year} {1992})}\BibitemShut {NoStop}%
\bibitem [{\citenamefont {Fr{\"o}wis}\ \emph {et~al.}(2015)\citenamefont
  {Fr{\"o}wis}, \citenamefont {Sangouard},\ and\ \citenamefont
  {Gisin}}]{frowis2015linking}%
  \BibitemOpen
  \bibfield  {author} {\bibinfo {author} {\bibfnamefont {F.}~\bibnamefont
  {Fr{\"o}wis}}, \bibinfo {author} {\bibfnamefont {N.}~\bibnamefont
  {Sangouard}},\ and\ \bibinfo {author} {\bibfnamefont {N.}~\bibnamefont
  {Gisin}},\ }\href@noop {} {\bibfield  {journal} {\bibinfo  {journal} {Optics
  Communications}\ }\textbf {\bibinfo {volume} {337}},\ \bibinfo {pages} {2}
  (\bibinfo {year} {2015})}\BibitemShut {NoStop}%
\bibitem [{\citenamefont {Streltsov}\ \emph {et~al.}(2017)\citenamefont
  {Streltsov}, \citenamefont {Adesso},\ and\ \citenamefont
  {Plenio}}]{streltsov2017colloquium}%
  \BibitemOpen
  \bibfield  {author} {\bibinfo {author} {\bibfnamefont {A.}~\bibnamefont
  {Streltsov}}, \bibinfo {author} {\bibfnamefont {G.}~\bibnamefont {Adesso}},\
  and\ \bibinfo {author} {\bibfnamefont {M.~B.}\ \bibnamefont {Plenio}},\
  }\href@noop {} {\bibfield  {journal} {\bibinfo  {journal} {Reviews of Modern
  Physics}\ }\textbf {\bibinfo {volume} {89}},\ \bibinfo {pages} {041003}
  (\bibinfo {year} {2017})}\BibitemShut {NoStop}%
\bibitem [{\citenamefont {Yadin}\ and\ \citenamefont
  {Vedral}(2016)}]{yadin2016general}%
  \BibitemOpen
  \bibfield  {author} {\bibinfo {author} {\bibfnamefont {B.}~\bibnamefont
  {Yadin}}\ and\ \bibinfo {author} {\bibfnamefont {V.}~\bibnamefont {Vedral}},\
  }\href@noop {} {\bibfield  {journal} {\bibinfo  {journal} {Physical Review
  A}\ }\textbf {\bibinfo {volume} {93}},\ \bibinfo {pages} {022122} (\bibinfo
  {year} {2016})}\BibitemShut {NoStop}%
\bibitem [{\citenamefont {Baumgratz}\ \emph {et~al.}(2014)\citenamefont
  {Baumgratz}, \citenamefont {Cramer},\ and\ \citenamefont
  {Plenio}}]{baumgratz2014quantifying}%
  \BibitemOpen
  \bibfield  {author} {\bibinfo {author} {\bibfnamefont {T.}~\bibnamefont
  {Baumgratz}}, \bibinfo {author} {\bibfnamefont {M.}~\bibnamefont {Cramer}},\
  and\ \bibinfo {author} {\bibfnamefont {M.~B.}\ \bibnamefont {Plenio}},\
  }\href@noop {} {\bibfield  {journal} {\bibinfo  {journal} {Physical review
  letters}\ }\textbf {\bibinfo {volume} {113}},\ \bibinfo {pages} {140401}
  (\bibinfo {year} {2014})}\BibitemShut {NoStop}%
\bibitem [{\citenamefont {Korsbakken}\ \emph {et~al.}(2007)\citenamefont
  {Korsbakken}, \citenamefont {Whaley}, \citenamefont {Dubois},\ and\
  \citenamefont {Cirac}}]{korsbakken2007measurement}%
  \BibitemOpen
  \bibfield  {author} {\bibinfo {author} {\bibfnamefont {J.~I.}\ \bibnamefont
  {Korsbakken}}, \bibinfo {author} {\bibfnamefont {K.~B.}\ \bibnamefont
  {Whaley}}, \bibinfo {author} {\bibfnamefont {J.}~\bibnamefont {Dubois}},\
  and\ \bibinfo {author} {\bibfnamefont {J.~I.}\ \bibnamefont {Cirac}},\
  }\href@noop {} {\bibfield  {journal} {\bibinfo  {journal} {Physical Review
  A}\ }\textbf {\bibinfo {volume} {75}},\ \bibinfo {pages} {042106} (\bibinfo
  {year} {2007})}\BibitemShut {NoStop}%
\bibitem [{\citenamefont {Dicke}(1954)}]{dicke1954coherence}%
  \BibitemOpen
  \bibfield  {author} {\bibinfo {author} {\bibfnamefont {R.~H.}\ \bibnamefont
  {Dicke}},\ }\href@noop {} {\bibfield  {journal} {\bibinfo  {journal}
  {Physical Review}\ }\textbf {\bibinfo {volume} {93}},\ \bibinfo {pages} {99}
  (\bibinfo {year} {1954})}\BibitemShut {NoStop}%
\bibitem [{\citenamefont {D{\"u}r}\ \emph {et~al.}(2000)\citenamefont
  {D{\"u}r}, \citenamefont {Vidal},\ and\ \citenamefont
  {Cirac}}]{dur2000three}%
  \BibitemOpen
  \bibfield  {author} {\bibinfo {author} {\bibfnamefont {W.}~\bibnamefont
  {D{\"u}r}}, \bibinfo {author} {\bibfnamefont {G.}~\bibnamefont {Vidal}},\
  and\ \bibinfo {author} {\bibfnamefont {J.~I.}\ \bibnamefont {Cirac}},\
  }\href@noop {} {\bibfield  {journal} {\bibinfo  {journal} {Physical Review
  A}\ }\textbf {\bibinfo {volume} {62}},\ \bibinfo {pages} {062314} (\bibinfo
  {year} {2000})}\BibitemShut {NoStop}%
\bibitem [{\citenamefont {Gerry}\ \emph {et~al.}(2005)\citenamefont {Gerry},
  \citenamefont {Knight},\ and\ \citenamefont
  {Knight}}]{gerry2005introductory}%
  \BibitemOpen
  \bibfield  {author} {\bibinfo {author} {\bibfnamefont {C.}~\bibnamefont
  {Gerry}}, \bibinfo {author} {\bibfnamefont {P.}~\bibnamefont {Knight}},\ and\
  \bibinfo {author} {\bibfnamefont {P.~L.}\ \bibnamefont {Knight}},\
  }\href@noop {} {\emph {\bibinfo {title} {Introductory quantum optics}}}\
  (\bibinfo  {publisher} {Cambridge university press},\ \bibinfo {year}
  {2005})\BibitemShut {NoStop}%
\end{thebibliography}%

 \appendix

\section{MMQS}
\label{SM-MMQS}
\textbf{Theorem:} In a system and in the basis of eigenvectors of the bounded observable $\hat{A}$ which does not have degeneracy, the state 
\begin{equation}
    |\psi_{\mathrm{MMQS}}\rangle=\frac{|a_{0}\rangle+e^{i\phi}|a_{N}\rangle}{\sqrt{2}}
    \label{MMQSProof}
\end{equation}
maximizes the measure $M$. Here,  $|a_{0}\rangle$ and $|a_{N}\rangle$ are the eigenvectors of $\hat{A}$ with minimum and maximum eigenvalues respectively and $\phi$ is a phase. Irrespective of $\phi$, $|\psi_{\mathrm{MMQS}}\rangle$ is unique.

\textbf{Proof:} First of all we prove the state $|\psi_{\mathrm{MMQS}}\rangle$ maximizes $M$ among all pure states in the range of spectrum of $\hat{A}$. Consider an arbitrary pure state $|\psi\rangle$ in the spectrum of $\hat{A}$ as below:
 \begin{equation}
|\psi\rangle=\sum_{i=0}^{N}c_{i}|a_{i}\rangle \equiv \rho=\sum_{ij=0}^{N}c_{i}c^{*}_{j}|a_{i}\rangle\langle a_{j}|,
\end{equation}
the $|a_{k}\rangle$s $k\in \{0,...,N\}$ are the eigenvectors of $\hat{A}$ corresponding to the eigenvalues $a_{k}$. If $i\leq j$ $i,j\in \{0,...,N \}$ then $a_{i}\leq a_{j}$. For $|\psi\rangle$ the measure is:
\begin{equation}
M=\frac{\sum_{ij}|c_{i}||c_{j}|d_{ij}}{\sum_{ij}|c_{i}||c_{j}|}. \label{ProofIMQS}
\end{equation}
We know that $\sum_{i=0}^{N}|c_{i}|^2=1$. Maximizing $M$, we neglect this constraint and at last we will turn back to it.

Differentiating $M$ in $|c_{k}|$ and equate it to zero, we find the below set of equations:
\begin{align}
\forall k\in\{0,...,N\},\frac{\partial M}{\partial |c_{k}|}=\frac{\partial}{\partial |c_{k}|}(\frac{\sum_{ij}|c_{i}||c_{j}|d_{ij}}{\sum_{ij}|c_{i}||c_{j}|})\notag &\\
=\frac{(2\sum_{i}|c_{i}|d_{ik})(\sum_{ij}|c_{i}||c_{j}|)}{(\sum_{ij}|c_{i}||c_{j}|)^2}\notag \\
-\frac{(2\sum_{i}|c_{i}|)(\sum_{ij}|c_{i}||c_{j}|d_{ij})}{(\sum_{ij}|c_{i}||c_{j}|)^2}=0.\label{IMQSPure}
\end{align}
As $M=\frac{\sum_{ij}|c_{i}||c_{j}|d_{ij}}{\sum_{ij}|c_{i}||c_{j}|}$, we substitute $M$ in the second fraction of the relation \ref{IMQSPure}, thus the equations \ref{IMQSPure} are simplified:
\begin{align}
\forall k\in\{0,...,N\},\frac{\sum_{i}|c_{i}|d_{ik}-\sum_{i}|c_{i}|M}{\sum_{ij}|c_{i}||c_{j}|}=0\notag \\
\Leftrightarrow M\sum_{i}|c_{i}|=\sum_{i}|c_{i}|d_{ik} \notag \\
\Leftrightarrow M=\frac{\sum_{i}|c_{i}|d_{ik}}{\sum_{i}|c_{i}|}.\label{IMQSPure1}
\end{align}
The $c_{k}$s maximizing $M$, satisfy the equations \ref{IMQSPure1}.

Now consider the equations associated with $k=0$ and $k=1$:
\begin{align}
k=0,M=\frac{\sum_{i}|c_{i}|d_{i0}}{\sum_{i}|c_{i}|},\notag & \\
k=1,M=\frac{\sum_{i}|c_{i}|d_{i1}}{\sum_{i}|c_{i}|}.\label{IMQSK}
\end{align}
By cross multiplication, we can write:
\begin{align}
k=1,M=\frac{\sum_{i}|c_{i}|d_{i1}}{\sum_{i}|c_{i}|} \Leftrightarrow M\sum_{i}|c_{i}|=\sum_{i}|c_{i}|d_{i1}\notag \\
\Leftrightarrow M\sum_{i}|c_{i}|=|c_{0}|d_{01}+\sum_{i\neq 0}|c_{i}|d_{i1}.\label{IMQSPure2}
\end{align}
Knowing $d_{i1}=d_{i0}-d_{10}$ for $i>1$ and replace it in \ref{IMQSPure2}:
\begin{align}
M\sum_{i}|c_{i}|=\sum_{i}|c_{i}|d_{i1}=|c_{0}|d_{10}+\sum_{i\neq 0}|c_{i}|(d_{i0}-d_{10}) \notag & \\
=(|c_{0}|-\sum_{i\neq 0}|c_{i}|)d_{10}+\sum_{i\neq 0}|c_{i}|d_{i0}.\label{C0-1}
\end{align}
Regarding the relations \ref{IMQSK}, the last term in \ref{C0-1} is  $M\sum_{i}|c_{i}|$ so they cancel each other and we have:
\begin{equation}
 |c_{0}|=|c_{N}|+\sum_{i\neq 0,N}|c_{i}|.\label{C01}
\end{equation}
Now we do the same procedure for $k=N$ and $k=N-1$,
\begin{align}
k=N-1,M=\frac{\sum_{i}|c_{i}|d_{i,N-1}}{\sum_{i}|c_{i}|},\notag & \\
k=N,M=\frac{\sum_{i}|c_{i}|d_{i,N}}{\sum_{i}|c_{i}|}.\label{IMQSKN}
\end{align}
By cross multiplication, we can write:
\begin{align}
k=N-1,M=\frac{\sum_{i}|c_{i}|d_{i,N-1}}{\sum_{i}|c_{i}|} \notag \\
\Leftrightarrow M\sum_{i}|c_{i}|=\sum_{i}|c_{i}|d_{i,N-1}\notag \\
\Leftrightarrow M\sum_{i}|c_{i}|=|c_{N}|d_{N,N-1}+\sum_{i\neq N}|c_{i}|d_{i,N-1}.\label{IMQSPure2N}
\end{align}
Knowing $d_{i,N-1}=d_{i,N}-d_{N,N-1}$ for $i<N-1$ and replace it in \ref{IMQSPure2N}:
\begin{align}
M\sum_{i}|c_{i}|=\sum_{i}|c_{i}|d_{i,N-1}\notag \\
=|c_{N}|d_{N,N-1}+\sum_{i\neq N}|c_{i}|(d_{i,N}-d_{N,N-1}) \notag & \\
=(|c_{N}|-\sum_{i\neq N}|c_{i}|)d_{N,N-1}+\sum_{i\neq N}|c_{i}|d_{i,N}. \label{C0-1N}
\end{align}
Regarding the relations \ref{IMQSKN}, the last term in \ref{C0-1N} is  $M\sum_{i}|c_{i}|$ so they cancel each other and we have:
\begin{equation}
 |c_{0}|=|c_{N}|-\sum_{i\neq 0,N}|c_{i}|.\label{C02}
\end{equation}
The equations \ref{C01} and \ref{C02} implies that:
\begin{align}
k\neq 0,N\rightarrow c_{i}=0 \notag &\\
|c_{0}|=|c_{N}|.
\end{align}
Hence, when $|c_{0}|=|c_{N}|$ and the other $c_{i}$s are zero, $M$ is extremum. If $|c_{0}|=|c_{N}|$ and also nonzero, the extremum is maximum too, because for all nonzero values of $|c_{0}|=|c_{N}|$ regardless of any constraints, the amount of extremum is $\frac{d_{max}}{2}$:
\begin{equation}
M_{max}=\frac{2 |c_{0}|^2\times 0 + 2 |c_{0}|^2\times d_{max}}{2|c_{0}|^2+2|c_{0}|^2}=\frac{d_{max}}{2}.
\end{equation}
the state $|\psi_{\mathrm{MMQS}}\rangle$ is the only pure state in the range of the spectrum of $\hat{A}$ that $|c_{0}|=|c_{N}|\neq 0$, therefore it maximizes $M$.

\textbf{Note:} Generally, by doing the exact same procedure for each $k$ and $k+1$, the set of equations in \ref{IMQSPure} turns to the below set of equations which are equivalent to  \ref{IMQSPure}:
\begin{equation}
\forall k\in\{0,...N\},\sum_{i=0}^{k}|c_{i}|^2=\sum_{i=k+1}^{N}|c_{i}|^2.
\end{equation} 
These equations only have answers when either all $c_{i}$s are zero (in this case $M=0$ and is minimum) or just $|c_{0}|$ and $|c_{N}|$ are nonzero and equal. The latter obtains the maximum for $M$.
 
Now, we prove the state \ref{MMQSProof}, also maximizes $M$ among mixed states in the spectrum of $\hat{A}$. 

Consider the mixed state $\rho$, we can decompose it in $N$ ensembles:
\begin{equation}
\rho=\sum_{i}^{N}P_{i}|\psi_{i}\rangle\langle\psi_{i}|, \label{Ensemble}
\end{equation}
which $i\in\{0,...,N\}$ and $|\psi_{i}\rangle$s are orthogonal. The $|\psi_{i}\rangle$s and their corresponding density matrices can be written as follow:
\begin{equation}
|\psi_{i}\rangle=\sum_{x=0}^{N}c_{x}^{i}|a_{x}\rangle \equiv |\psi_{i}\rangle\langle \psi_{i}|=\sum_{x,y=0}^{N}a_{xy}^{i}|a_{x}\rangle\langle a_{y}|,
\end{equation}
which $x,y\in\{0,...,N\}$. We know the relations between $a_{xy}^{i}$ and $c_{x}^{i}$:
\begin{align}
a_{xy}^{i}=c_{x}^{i}c_{y}^{i*},\notag \\
a_{xx}^{i}=c_{x}^{i}c_{x}^{i*}=|c_{x}|^2.
\end{align}
Also,
\begin{equation}
|c_{y}|^2=\frac{|c_{x}^{i}|^2|c_{y}^{i}|^2}{|c_{x}^{i}|^2}=\frac{|a_{xy}^{i}|^2}{a_{xx}}.\label{PreConstraint}
\end{equation}
Regarding $\sum_{y}|c_{y}|^2=1$ and with respect to the equation \ref{PreConstraint}, we have the following constraints for $a_{xy}^{i}$:
\begin{align}
\sum_{y}|c_{y}|^2=1\Rightarrow \sum_{y}\frac{|a_{xy}^{i}|^2}{a_{xx}}=1\notag \\
f_{x}^{i}=\sum_{y}|a_{xy}^{i}|^2-a_{xx}^{i}=0 \label{Constraint1}
\end{align}
and
\begin{align}
\sum_{x}|c_{x}|^2=1 \Rightarrow \sum_{x}a_{xx}=1 \notag \\
f_{0}^{i}=\sum_{x}a_{xx}=1.\label{Constraint2}
\end{align}
The measure $M$ for $\rho$ is:
\begin{align}
M=\frac{\sum_{x,y}|\sum_{i}P_{i}a_{xy}^{i}|d_{xy}}{\sum_{x,y}|\sum_{i}P_{i}a_{xy}^{i}|}\notag \\
=\frac{\sum_{x,y}\sqrt{(\sum_{i}P_{i}a_{xyR}^{i})^2+(\sum_{i}P_{i}a_{xyI}^{i})^2}d_{xy}}{\sum_{x,y}\sqrt{(\sum_{i}P_{i}a_{xyR}^{i})^2+(\sum_{i}P_{i}a_{xyI}^{i})^2}}.\label{IMQSMix}
\end{align}
$a_{xyR}^{i}$ and $a_{xyI}^{i}$ are the real and imaginary parts of $a_{xy}^{i}$ respectively. Besides, we denote the denominator of $M$ in the right side of \ref{IMQSMix} with $D$.

Maximizing $M$, we differentiate $M$ in $a_{xyR}^{i}$ and $a_{xyI}^{i}$ and with respect to the constraints $f_{x}^{i}$, we use Lagrange multipliers method. We can directly apply the constraints \ref{Constraint2} in $D$:
\begin{align}
D=\sum_{x,y}\sqrt{(\sum_{i}P_{i}a_{xyR}^{i})^2+(\sum_{i}P_{i}a_{xyI}^{i})^2}\notag \\ =\sum_{x}\sqrt{(\sum_{i}P_{i}a_{xxR}^{i})^2+(\sum_{i}P_{i}a_{xxI}^{i})^2}\notag\\
+\sum_{x,y,x\neq y}\sqrt{(\sum_{i}P_{i}a_{xyR}^{i})^2+(\sum_{i}P_{i}a_{xyI}^{i})^2},
\end{align}
because $a_{xx}^{i}\geq 0$ are real, we can write:
\begin{align}
D=\sum_{i}P_{i}\sum_{x}a_{xx}^{i}\notag \\
+\sum_{x,y,x\neq y}\sqrt{(\sum_{i}P_{i}a_{xyR}^{i})^2
+(\sum_{i}P_{i}a_{xyI}^{i})^2}.\label{SimpleDenom}
\end{align}
By the constraints  \ref{Constraint2}, the first term in the right side of \ref{SimpleDenom} is $\sum_{i}P_{i}\sum_{x}a_{xx}^{i}=1$, thus
\begin{equation}
D=1+\sum_{x,y,x\neq y}\sqrt{(\sum_{i}P_{i}a_{xyR}^{i})^2+(\sum_{i}P_{i}a_{xyI}^{i})^2}.
\end{equation}
\textbf{Note:} By applying the constraints \ref{Constraint2} and with respect to $d_{xx}=0$, $M$ is no longer a function of $a_{xy}^{i}$.

Carrying the calculations for maximizing $M$, we reach to the equations below:
\begin{align}
x\neq y,\frac{\rho_{xyR}(d_{xy}-M)}{D|\rho_{xy}|}=\frac{\lambda_{x}^{i}}{2P_{i}}a_{xyR}^{i}, \label{1}\\
x\neq y,\frac{\rho_{xyI}(d_{xy}-M)}{D|\rho_{xy}|}=\frac{\lambda_{x}^{i}}{2P_{i}}a_{xyI}^{i}, \label{2}\\
\lambda_{x}^{i}(2a_{xx}^{i}-1)=0.\label{3} 
\end{align}
$\rho_{xyR}=Re(\rho_{xy})$, $\rho_{xyI}=Im(\rho_{xy})$ and $\lambda_{x}^{i}$s are the Lagrange multipliers associated with 
$f_{x}^{i}$.

We show the calculations for deriving the equations \ref{1}; The other equations are derived in the same way. By differentiating $M$ in $a_{xyR}^{i}$,
\begin{align}
\frac{\partial M}{\partial a_{xyR}^{i}}=\frac{\frac{4P_{i}(\sum_{i}P_{i}a_{xyR}^{i})d_{xy}}{\sqrt{(\sum_{i}P_{i}a_{xyR}^{i})^2+(\sum_{i}P_{i}a_{xyI}^{i})^2}}D}{D^2}\notag \\
-\frac{\frac{4P_{i}(\sum_{i}P_{i}a_{xyR}^{i})}{\sqrt{(\sum_{i}P_{i}a_{xyR}^{i})^2+(\sum_{i}P_{i}a_{xyI}^{i})^2}}}{D^2}\notag\\
\times \sum_{x,y}\sqrt{(\sum_{i}P_{i}a_{xyR}^{i})^2+(\sum_{i}P_{i}a_{xyI}^{i})^2}d_{xy}.\label{SampleDerive}
\end{align}
In the above relation we can substitute the following terms:
\begin{align}
\rho_{xy}=\sqrt{(\sum_{i}P_{i}a_{xyR}^{i})^2+(\sum_{i}P_{i}a_{xyI}^{i})^2},
\end{align}
\begin{equation}
\rho_{xyR}=\sum_{i}P_{i}a_{xyR}^{i},
\end{equation}
\begin{equation}
M=\frac{\sum_{x,y}\sqrt{(\sum_{i}P_{i}a_{xyR}^{i})^2+(\sum_{i}P_{i}a_{xyI}^{i})^2}d_{xy}}{D}.
\end{equation}
With these substitutions, $\frac{\partial M}{\partial a_{xyR}^{i}}$ in \ref{SampleDerive} is simplified as:
\begin{equation}
\frac{\partial M}{\partial a_{xyR}^{i}}=\frac{4P_{i}\rho_{xyR}(d_{xy}-M)}{D|\rho_{xy}|}
\end{equation}
Applying the constraints \ref{Constraint1} is by subtracting $\lambda_{x}^{i}\frac{\partial f_{x}^{i}}{\partial a_{xyR}^{i}}$ from $\frac{\partial M}{\partial a_{xyR}^{i}}$. Because
\begin{equation}
\frac{\partial f_{x}^{i}}{\partial a_{xyR}^{i}}=\frac{\partial}{\partial a_{xyR}^{i}}(\sum_{y}|a_{xy}^{i}|^2-a_{xx}^{i})=2a_{xyR}^{i},
\end{equation}
at last we end up the following equation:
\begin{equation*}
\frac{\rho_{xyR}(d_{xy}-M)}{D|\rho_{xy}|}=\frac{\lambda_{x}^{i}}{2P_{i}}a_{xyR}^{i}
\end{equation*}
which is the same with \ref{1}.

Replacing $a_{xyR}^{i}$ and $a_{xyI}^{i}$ from equations \ref{1} and\ref{2} in \ref{Constraint1},
\begin{equation}
\sum_{y\neq x}\frac{((\rho_{xyR})^2+(\rho_{xyI})^2)(d_{xy}-M)^2}{D^2|\rho_{xy}|^2}=\frac{\lambda_{x}^{i2}}{4P_{i}^2}(a_{xx}^{i2}-a_{xx}^{i}).
\end{equation}
Because $\rho_{xyR}^{2}+\rho_{xyI}^{2}=|\rho_{xy}|^2$,
\begin{equation}
\sum_{y\neq x}\frac{(d_{xy}-M)^2}{D^2}=\frac{\lambda_{x}^{i2}}{4P_{i}^2}(a_{xx}^{i2}-a_{xx}^{i}).\label{LambdaNotZero}
\end{equation}
Since $d_{xy}$s have different amounts (because $\hat{A}$ does not have degeneracy) and $M\geq\frac{d_{max}}{2}$ (if $M\leq\frac{d_{max}}{2}$ the theorem is proven because $M(|\psi_{MMQS}\rangle)=\frac{d_{max}}{2}$), the left side of \ref{LambdaNotZero} is positive, therefore $\lambda_{x}^{i}$s must be nonzero.
 
From the equations \ref{3}, $a_{xx}^{i}$s are $\frac{1}{2}$, consequently from the constraints \ref{Constraint2} we find that $|\psi_{i}\rangle$s are pure states with $2\times 2$ density matrices in which the diagonal elements are $\frac{1}{2}$, therefore $|\psi_{i}\rangle$s must be of the form $(e^{i\phi}|a_{i}\rangle+|a_{j}\rangle)/\sqrt{2}$. Thus if only all $|\psi_{i}\rangle=|\psi_{\mathrm{MMQS}}\rangle$, $M$ is maximized and $\rho$ is the density matrix corresponding to $|\psi_{\mathrm{MMQS}}\rangle$ and the theorem is proved.

 \section{Calculation of the Measure for the Uniform State}
 \label{UniformCalculation}
Here we calculate the measure for Uniform state in the basis of total spin-z .Total spin-z in a spin ensemble system in which the particles take the values $0$ or $1$ for the spin-z observable, is equal to the number of particles having the value of spin-z equal to $1$, so in the basis of total spin-z we can represent the density matrix of the Uniform state as below:
 \begin{equation}
  \frac{1}{2^{N}}   \sum_{\{i\},\{j\}} |i_{1}i_{2}...i_{N}\rangle\langle j_{1}j_{2}...j_{N}| 
 \end{equation}
 in which $i_{k}$ and $j_{k}$ are indicating the spin of the $k$'th particle in z-direction and takes the values $0$ or $1$.

First we calculate $P_{(d)}$, we need to find the density matrix elements corresponding to the distances with amount of $d$. These elements are those in which the discrepancy of numbers of $1$ in $|i_{1}i_{2}...i_{N}\rangle$ and $\langle j_{1}j_{2}...j_{N}|$ is equal to $d$. If $|i_{1}i_{2}...i_{N}\rangle$ has total z-magnetization equal to $m$, then $m$ number of $i_{k}$s must be $1$ and the others ($N-m$) are zero, so we have $\binom{N}{m}$ possible choices, in order that we require $|i_{1}i_{2}...i_{N}\rangle\langle j_{1}j_{2}...j_{N}|$ to be associated with the distance $d$, the total z-magnetization of $\langle j_{1}j_{2}...j_{N}|$ must be $m+d$ or $m-d$, that for the first we have $\binom{N}{m+d}$ and for the last we have $\binom{N}{m-d}$ possible choices; Thus based on the product rule, the number of elements associated with the distance $d$ is:
 \begin{equation}
     N_{d}=2\sum_{m=0}^{N} \binom{N}{m} \binom{N}{m+d}.
 \end{equation}
 Because all of the elements in the density matrix of Uniform state are equal to $\frac{1}{2^{N}}$, and the number of elements is $2^N\times2^{N}$,
  \begin{equation}
    P_{(d)}=\frac{N_{d}} {2^{N}*2^{N}}=\frac{2\sum_{m=0}^{N} \binom{N}{m}\binom{N}{m+d}}{2^{2N}}.
 \end{equation}

Having $P_{(d)}$, we can calculate the measure directly for this state:
\begin{equation}
    M=\frac{2\sum_{d=0}^{N} \sum_{m=0}^{N} \binom{N}{m} \binom{N}{m+d} d}{2^{2N}}. \label{USMeasureDirectly}
\end{equation}
$M$ can be simplified as below:
\begin{equation}
    M=\frac{(N+1)!(2N+1)!}{N!(N+2)! 2^{2N}}
\end{equation}
At last, in the limit $N>>1$ using Stirling approximation, we have:
\begin{equation*}
    M=e^{N ln (\frac{(N+\frac{1}{2})^2}{(N-1)(N+2)})}
\end{equation*}

\section{Generalized GHZ}
Another interesting state is the the generalized GHZ considered in \cite{dur2002effective,frowis2018macroscopic}. 
The generalized GHZ state 
is defined as
\begin{equation}
  |\phi_{\epsilon}\rangle=\frac{|0\rangle^{\otimes N} + (\cos{\epsilon}|0 \rangle+ \sin{\epsilon} |1 \rangle)^{\otimes N}}{2+2\cos^{N}\epsilon} \label{GGHZ}.
\end{equation}

We calculate the measure for this state in the limits $N>>1$  and $\epsilon <<1$ and $N\epsilon<1$ which gives
\begin{equation}
    M_{\mathrm{GHZ}_{\epsilon}}\approx\frac{N\epsilon}{2}. \label{MeasureGGHZ}
\end{equation}

As we see, quantum macroscopicity of generalized GHZ, evaluated by our measure, is plausible compared to the amount that the D\"{u}r et al. measures obtain in the same limits; which is $N\epsilon^{2}$ \cite{dur2000three}.

\section{Some Other Photonic States}
\subsection{Superposition of Coherent States(SCS) \cite{buvzek1992superpositions}}
 SCS is the superposition of two coherent states with annihilation operator eigenvalues of $\alpha$ and $ - \alpha$. 
It is defined as below:
 \begin{equation}
     |\mathrm{SCS}\rangle  = ( |\alpha \rangle+|-\alpha \rangle)/z \label{SCS}
 \end{equation}
where $z=\sqrt{2+2Re(\langle\alpha|-\alpha\rangle)}$ is the normalization factor. 

In the coherent state $|\alpha\rangle$ consist in large numbers of photons, the amount of $|\alpha|$ is also large and for large amounts of $|\alpha|$ we can consider $|\alpha\rangle$ and $|-\alpha\rangle$ orthogonal to each other(i.e. $\langle \alpha|-\alpha\rangle \approx 0$) \cite{gerry2005introductory}, therefore, in the basis of the quadrature $X\cos{\theta}+P\sin{\theta}$ with $\tan\theta=\frac{Im(\alpha)}{Re(\alpha)}$,  the density matrix can be approximated as below for large $|\alpha|$:
\begin{align}
    \rho_{\mathrm{SCS}}=\frac{1}{2}(|\alpha\rangle \langle \alpha |+|-\alpha\rangle \langle- \alpha |\notag \\
    |\alpha\rangle \langle -\alpha |+|-\alpha\rangle \langle \alpha |).\label{SCSApp}
\end{align}
As we see in \ref{SCSApp}, $P_{(d)}$ is distributed with the same probability of $\frac{1}{2}$ on the distances $0$ an $|\alpha-(-\alpha))|=2|\alpha|$ so in the aforementioned quadrature's basis, the measure is obtained as follow:
 \begin{equation}
       M_{SCS}\approx\frac{1}{2}\times 0+\frac{1}{2}\times 2|\alpha|=|\alpha|. \label{MeasureSCS}
 \end{equation}
Since $|\alpha|^2$ is the mean number of photons in the system, the measure increases by increasing the number of photons.
\subsection{Mixed SCS } 
Mixed SCS is defined as below:
\begin{equation}
    \rho \textrm{  } \propto \textrm{  } | \alpha \rangle\langle \alpha|+|-\alpha \rangle\langle-\alpha| \label{MixedSCS}
\end{equation}

In the case of $|\alpha|>>1$, the two coherent states $|\alpha \rangle$ and $|-\alpha \rangle$ could be considered orthogonal to each other \cite{gerry2005introductory}. Hence In the basis of the quadrature $X\cos{\theta}+P\sin{\theta}$ with $\tan\theta=\frac{Im(\alpha)}{Re(\alpha)}$ and for large amounts of $|\alpha|$, the density matrix of mixed SCS turns to a diagonal one and the measure becomes zero for the state. This result has meaning when we compare the mixed SCS with SCS; Compared to SCS, a mixed SCS has lost its coherence terms in the aforementioned basis and it should not be macroscopic quantum.

\subsection{Thermal State}
Thermal state\cite{gerry2005introductory} is a thermal classical mix of photons with the density matrix
\begin{equation}
    \rho_{\mathrm{Thermal}}=\frac{\sum_{N} e^{-\beta N}|N\rangle\langle N|}{Z}, \label{Thermal}
\end{equation}
$Z=\sum_{N}e^{-\beta N}$ is the normalization factor(i.e. in terms of statistical mechanics it is the partition function).
Because the density matrix has no coherence(off-diagonal) terms,
\begin{equation}
    M_{\mathrm{Thermal}}=0 \label{MeasureThermal}.
\end{equation}

\end{document}